\documentclass[aps,prb,twocolumn,superscriptaddress,floatfix,times]{revtex4}
\usepackage{graphicx}
\usepackage[dvips,unicode,colorlinks,linkcolor=blue,citecolor=blue,urlcolor=blue]{hyperref}
\begin{document}

\title{Vibrational Tamm states at the edges of graphene nanoribbons}

\author{Alexander V. Savin$^{1,2}$ and Yuri S. Kivshar}

\affiliation{Nonlinear Physics Center, Research School of Physics and Engineering,
Australian National University, Canberra, ACT 0200, Australia\\
$^2$Semenov Institute of Chemical Physics, Russian Academy
of Sciences, Moscow 119991, Russia}

\begin{abstract}
We study vibrational states localized at the edges of graphene nanoribbons.
Such surface oscillations can be considered as a phonon analog of Tamm states well known
in the electronic theory. We consider both armchair and zigzag graphene stripes and demonstrate
that surface modes correspond to phonons localized at the edges of the graphene nanoribbon,
and they can be classified as in-plane and out-of-plane modes. In addition, in armchair nanoribbons
anharmonic edge modes can experience longitudinal localization in the form of
self-localized nonlinear modes, or surface breather solitons.
\end{abstract}


\maketitle

\section{Introduction}

Graphite nanocrystallites are recently discovered carbon-based nanomaterials~\cite{p1}
which are a few atoms thick and stable under ambient conditions, and
they keep many promises as materials for nanoscale transistors~\cite{p2}. Remarkable properties of graphite
structures make their study one of the hot topics of nanoscience~\cite{p3}.

Graphene nanoribbons are effectively low-dimensional nanostructures similar to carbon nanotubes,
but they differ from nanotubes by the presence of edges. Due to the edges,
graphene nanoribbons can demonstrate many novel geometry-driven properties, depending on
their width and helicity. A majority of the current studies of graphene nanoribbons
target their electronic and magnetic properties stipulated by the presence of edges,
including the existence of the edge modes~\cite{lee}, which
are an analog of {\em surface states} in this two-dimensional geometry.

In general, surface states are spatially localized modes which can be generated at
surfaces~\cite{book_S}, and such surface states have been studied in many branches of
of physics including electrons in crystals~\cite{Tamm,Shockley},
surface phonons~\cite{Maradudin}, surface polaritons~\cite{Agranovich}, and optical
surface modes in waveguide arrays~\cite{ivan,ivan_exp}. In this paper, we analyze the
specific properties of vibrational phonon modes localized at the edges of graphene nanoribbons,
the so-called {\em edge phonon modes}. Such localized surface modes can be considered
as a phonon analog of Tamm states well known in the electronic theory~\cite{Tamm}.

Vibrational edge modes in graphene nanoribbons have been discussed recently
for the hydrogen-terminated graphene nanoribbons~\cite{p4}. Using the second-generation
reactive empirical bond-order potential and density-functional theory calculations, it was
demonstrated that two different edge modes can exist in armchair nanoribbons, whereas
the zigzag nanoribbons can support only one such mode. Importantly, all those edge
modes were found to correspond to the out-of-plane vibrations of the edge hydrogen atoms.

In this paper, we employ numerical analysis and molecular dynamics simulations to study
localized modes in graphene nanoribbons. We demonstrate that such nanostructures can support
much larger number of vibrational phonon modes localized at the edges.
More specifically, we find that the armchair nanoribbon can support 7 edge modes, whereas
the zigzag nanoribbon can support 12 edge modes. Only one part of such modes correspond
to out-of-plane oscillations of the edge atoms, whereas and the second part is
characterized by in-plane edge oscillations. Chemical modification of the edge
atoms can lead to the generation of larger number of localized surface modes.
In addition, numerical simulations of anharmonic oscillations demonstrate that in
armchair nanoribbons the vibrational energy can become self-trapped along the edge
of the stripe with the formation of both in-plane and out-of-plane surface breather solitons.

The paper is organized as follows. In Sec.~II we discuss our model including
the potentials for describing the interatomic interactions and the edge effects.
Section~III is devoted to a general linear analysis, and it discusses
the dispersion curves for both armchair and zigzag nanoribbons.
In Sec.~IV we describe the specific properties of the edge modes and corresponding energy
localization. Some properties of anharmonic oscillations and surface breather solitons
are discussed in Sec.~V, whereas Sec.~VI concludes the paper.

\section{Model of graphene nanoribbons}

\begin{figure}[htbp]
\begin{center}
\includegraphics[angle=0, width=1\linewidth]{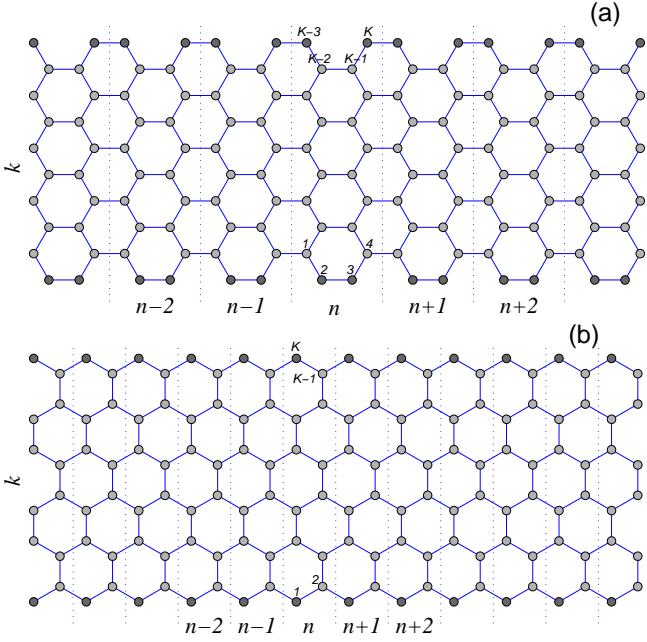}
\end{center}
\caption{\label{fg01}\protect
Schematic view of (a) armchair and (b) zigzag nanoribbons
with the atom numbering. The edge atoms are shown as filled circles.
Dotted lines separate the elementary cells of the nanoribbon. $M$ is the
number of atoms in the elementary cell.
}
\end{figure}

We model the graphene nanoribbon as a planar stripe of graphite, with the properties
depending on the stripe width
and chirality. Here we consider two most typical geometries of the armchair nanoribbon
[see Fig. \ref{fg01}(a)] and zigzag nanoribbon [see Fig. \ref{fg01}(b)].

The structure of the nanoribbon can be presented through a longitudinal repetition
of the elementary cell composed of $K$ atoms, as shown in Figs.~\ref{fg01}(a,b).
For the armchair, the number $K$ is multiple of 4, whereas for the zigzag nanoribbon
the number $K$ is multiple of 2. We use the atom numbering shown in Figs.~\ref{fg01}(a,b).
In this case, each carbon atom has a two-component index $\alpha=(n,k)$, where
$n=0,\pm1,\pm2,...$ stands for the number of the elementary cells, and $k=1,2,...,K$
stands for the number of atoms in the cell.

Each elementary cell of the armchair nanoribbon has four edge atoms, whereas the zigzag nanoribbon
has only two edge atoms. In Figures~\ref{fg01}(a,b), we show these edge atoms as filled circles.
In a realistic case, the edge atoms are chemically modified and, generally speaking,
the following situations are possible~\cite{p5,p6}: a single hydrogen CH
(hydrogen-terminated graphene nanoribbon), CH$_2$ -- two hydrogens, COH -- a hydroxyl group,
and CHOH -- a hydrogen atom and a hydroxyl group that
resulted from the water decomposition.

In this study, we do not take into account the specific chemical nature of the surface
atoms, but consider a change of the effective mass of the edge atom.
Therefore, in our model of graphene nanoribbons we take the mass of atoms inside the
stripe as $M_0=12m_p$, and for the edge atoms
we consider a larger mass, such as $M_1=13m_p$, $14m_p$, $29m_p$, $30m_p$ (where
$m_p=1.6603\cdot10^{-27}$kg is the proton mass).

\begin{figure}[t]
\begin{center}
\includegraphics[angle=0, width=1\linewidth]{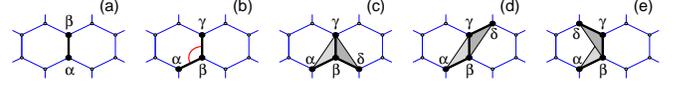}
\end{center}
\caption{\label{fg02}\protect
Configurations containing up to $i$-th nearest-neighbor interactions for:
(a) $i=1$, (b) $i=2$, (c) $i=3$, (d) $i=4$, and (e) $i=5$.
}
\end{figure}

To describe nanoribbon oscillations, we write the system Hamiltonian in the
form,
\begin{equation}
H=\sum_{n=-\infty}^{+\infty}\sum_{k=1}^K \left[\frac12 M_{(n,k)}
(\dot{\bf u}_{(n,k)},\dot{\bf u}_{(n,k)})+P_{(n,k)}\right],
\label{f1}
\end{equation}
where $M_\alpha$ is the mass of the hydrogen atom with the index $\alpha=(n,k)$
(for internal atoms we take $M_\alpha=M_0$, whereas for the edge atoms we take a larger mass,
$M_\alpha=M_1>M_0$), ${\bf u}_\alpha=(x_\alpha(t),y_\alpha(t),z_\alpha(t))$
is the radius-vector of the carbon atom with the index $\alpha$ at the moment $t$.
The term $P_\alpha$ describes the interaction of the atom with the index $\alpha=(n,k)$
and its neighboring atoms.
The potential depends on variations of bond length, bond angles, and dihedral angles
between the planes formed by three neighboring carbon atoms, and it can be written in the form
\begin{equation}
P=\sum_{\Omega_1} U_1+\sum_{\Omega_2} U_2+\sum_{\Omega_3} U_3+\sum_{\Omega_4} U_4
+\sum_{\Omega_5} U_5,
\label{f2}
\end{equation}
where $\Omega_i$, with $i=1,2,3,4,5$, are the sets of configurations including up to
nearest-neighbor interactions. Owing to a large redundancy, the sets only
need to contain configurations of the atoms shown in Fig. \ref{fg02}, including their
rotated and mirrored versions.

The potential $U_1({\bf u}_\alpha,{\bf u}_\beta)$ describes the deformation energy
due to a direct interaction between pairs of atoms with the indexes
$\alpha$ and $\beta$, as shown in Fig. \ref{fg02}(a).
The potential $U_2({\bf u}_\alpha,{\bf u}_\beta,{\bf u}_\gamma)$
describes the deformation energy of the angle between the valent bonds
${\bf u}_\alpha{\bf u}_\beta$ and  ${\bf u}_\beta{\bf u}_\gamma$, see Fig.~\ref{fg02}(b).
Potentials $U_{i}({\bf u}_\alpha,{\bf u}_\beta,{\bf u}_\gamma,{\bf u}_\delta)$, $i=3$, 4, 5,
describes the deformation energy associated with a change of the effective angle between
the planes ${\bf u}_\alpha,{\bf u}_\beta,{\bf u}_\gamma$ and
${\bf u}_\beta,{\bf u}_\gamma,{\bf u}_\delta$, as shown in Figs. \ref{fg02}(c,d,e).

We use the potentials employed in the modeling of the dynamics of large polymer
macromolecules \cite{p7,p8,p9,p10,p11} for the valent bond coupling,
\begin{equation}
U_1({\bf u}_1,{\bf u}_2)=\epsilon_1
\{\exp[-\alpha_0(\rho-\rho_0)]-1\}^2,~~~\rho=|{\bf u}_2-{\bf u}_1|,
\label{f3}
\end{equation}
where $\epsilon_1=4.9632$~eV is the energy of the valent bond and $\rho_0=1.418$~\AA~
is the equilibrium length of the bond;
the potential of the valent angle
\begin{eqnarray}
U_2({\bf u}_1,{\bf u}_2,{\bf u}_3)=\epsilon_2(\cos\varphi-\cos\varphi_0)^2,
\label{f4}\\
\cos\varphi=({\bf u}_3-{\bf u}_2,{\bf u}_1-{\bf u}_2)/
(|{\bf u}_3-{\bf u}_2|\cdot |{\bf u}_2-{\bf u}_1|),
\nonumber
\end{eqnarray}
so that the equilibrium value of the angle is defined as $\cos\varphi_0=\cos(2\pi/3)=-1/2$;
the potential of the torsion angle
\begin{eqnarray}
\label{f5}
U_i({\bf u}_1,{\bf u}_2,{\bf u}_3,{\bf u}_4)=\epsilon_i(1-z_i\cos\phi),\\
\cos\phi=({\bf v}_1,{\bf v}_2)/(|{\bf v}_1|\cdot |{\bf v}_2|),\nonumber \\
{\bf v}_1=({\bf u}_2-{\bf u}_1)\times ({\bf u}_3-{\bf u}_2), \nonumber \\
{\bf v}_2=({\bf u}_3-{\bf u}_2)\times ({\bf u}_3-{\bf u}_4), \nonumber
\end{eqnarray}
where the sign $z_i=1$ for the indices $i=3,4$ (equilibrium value of the torsional angle $\phi_0=0$)
and $z_i=-1$ for the index $i=5$ ($\phi_0=\pi$).

The specific values of the parameters are $\alpha_0=1.7889$~\AA$^{-1}$,
$\epsilon_2=1.3143$ eV, and $\epsilon_3=0.499$ eV, and they are found from the frequency
spectrum of small-amplitude oscillations of a sheet of graphite~\cite{p12}.
According to the results of Ref.~\cite{p13} the energy $\epsilon_4$ is close to
the energy $\epsilon_3$, whereas  $\epsilon_5\ll \epsilon_4$
($|\epsilon_5/\epsilon_4|<1/20$). Therefore, in what follows we use the values
$\epsilon_4=\epsilon_3=0.499$ eV and assume $\epsilon_5=0$, the latter means that we
omit the last term in the sum (\ref{f2}).

\section{Dispersion curves}

In the equilibrium state, $\{{\bf u}^0_{(n,k)}\}_{n=-\infty,~k=1}^{+\infty,~K}$,
all atoms are in the plane of the nanoribbon, so that all valent bonds have the
equilibrium length $\rho=\rho_0$, and all angle between the bonds coincide, $\varphi=2\pi/3$.
Armchair nanoribbon is characterized by the longitudinal shift $h_x=3\rho_0$ ¨
and width $D_y=\sqrt{3}(K/2-1)\rho_0/2$, where $K$ is the number of atoms in the
elementary cell.
Zigzag nanoribbon is characterized by the longitudinal shift $h_x=\sqrt{3}\rho_0$ and width
$D_y=\rho_0(3K-4)/4$.

Next, we introduce $3K$-dimensional vector,
$$
{\bf x_n}=({\bf u}_{(n,1)}-{\bf u}^0_{(n,1)},...,{\bf u}_{(n,K)}-{\bf u}^0_{(n,K)})
$$
describing a shift of the atom of the $n$-th cell from its equilibrium position. Then,
the nanoribbon Hamiltonian can be written in the following form,
\begin{equation}
H=\sum_n\{\frac12({\bf M}\dot{\bf x}_n,\dot{\bf x}_n)
+{\cal P}({\bf x}_{n-1},{\bf x}_{n},{\bf x}_{n+1})\},
\label{f6}
\end{equation}
where ${\bf M}$ is the diagonal matrix of masses of all atoms of the elementary cell.

Hamiltonian (\ref{f6}) generates the following set of the equations
of motion:
\begin{eqnarray}
-{\bf M}\ddot{\bf x}_n&=&{\cal P}_{{\bf x}_1}({\bf x}_{n},{\bf x}_{n+1},{\bf x}_{n+2})
+{\cal P}_{{\bf x}_2}({\bf x}_{n-1},{\bf x}_{n},{\bf x}_{n+1})\nonumber\\
&&+{\cal P}_{{\bf x}_3}({\bf x}_{n-2},{\bf x}_{n-1},{\bf x}_{n})
\label{f7}
\end{eqnarray}
In the linear approximation, this system takes the form
\begin{equation}
-{\bf M}\ddot{\bf x}_n={\bf B}_1{\bf x}_n+{\bf B}_2{\bf x}_{n+1}+{\bf B}^*_2{\bf x}_{n-1}
+{\bf B}_3{\bf x}_{n+2}+{\bf B}^*_3{\bf x}_{n-2},
\label{f8}
\end{equation}
where the matrix elements are defined as
\begin{eqnarray}
{\bf B}_1={\cal P}_{{\bf x}_1{\bf x}_1}
         +{\cal P}_{{\bf x}_2{\bf x}_2}
         +{\cal P}_{{\bf x}_3{\bf x}_3},\nonumber \\
{\bf B}_2={\cal P}_{{\bf x}_1{\bf x}_2}+{\cal P}_{{\bf x}_2{\bf x}_3},~
{\bf B}_3={\cal P}_{{\bf x}_1{\bf x}_3}, \nonumber
\end{eqnarray}
and the matrix of the partial derivatives takes the form
$$
{\cal P}_{{\bf x}_i{\bf x}_j}=\frac{\partial^2\cal P}{\partial{\bf x}_i\partial{\bf x}_j}
({\bf 0},{\bf 0},{\bf 0}),~~i,j=1,2,3.
$$
Solutions of the system of linear equations (\ref{f8}) can
be sought in the standard form,
\begin{equation}
{\bf x}_n=A{\bf e}\exp(iqn-i\omega t),
\label{f9}
\end{equation}
where $A$ is the mode amplitude, ${\bf e}$ is the normalized dimensionless vector
($({\bf Me},{\bf e})/M_0=1)$, $q\in[0,\pi]$ is the dimensionless wave number, and
$\omega$ is the phonon frequency.
Substituting the expression (\ref{f9}) into the system (\ref{f8}), we obtain
the eigenvalue problem
\begin{equation}
\omega^2{\bf Me}=[{\bf B}_1+{\bf B}_2e^{iq}+{\bf B}_2^*e^{-iq}+{\bf B}_3e^{2iq}
+{\bf B}_3^*e^{-2iq}]{\bf e}.
\label{f10}
\end{equation}

\begin{figure}[t]
\includegraphics[angle=0, width=1.\linewidth]{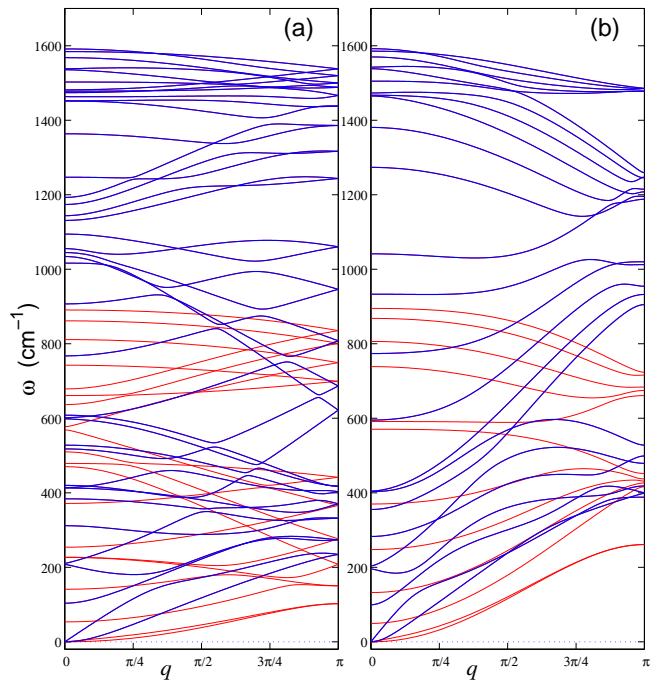}
\caption{ (color online)
Structure of $3K$ dispersion curves for (a) armchair nanoribbon with the width $D_y=11.05$~\AA~
($K=20$), and (b) zigzag nanoribbon with the width $D_y=11.34$~\AA~ ($K=12$).
Blue (black) curves correspond to the in-plane vibrations, and red (gray) curves correspond
to out-of-plane vibrations. Mass of the edge atoms is $M_1=13m_p$.
}
\label{fg03}
\end{figure}

Therefore, in order to find the dispersion relations characterizing
the modes of the nanoribbon for each fixed value of the
wave number $0\le q\le\pi$  we need to find the eigenvalues of the
Hermitian matrix (\ref{f10}) of the order $3K\times 3K$. As a result,
the dispersion curves are composed of $3K$ branches $\{\omega_j(q)\}_{j=1}^{3K}$,
as shown in Fig. \ref{fg03}.  Two thirds of the branches corresponds to the atom vibrations
in the plane of the nanoribbon $xy$ (in-plane vibrations), whereas only one third
corresponds to the vibrations orthogonal to the plane (out-of-plane vibrations),
when the atoms are shifted along the axes $z$.

As is seen in Fig. \ref{fg03}, the spectrum of the nanoribbon oscillations
occupies the frequency interval $[0,\omega_m]$, where the
maximum (cutoff) frequency $\omega_m$ practically does not depend
on the chirality. The maximum frequency $\omega_m=1600$~cm$^{-1}$. This value agrees well
with the experimental data for a planar graphite \cite{p14,p15}.
Four branches of the dispersion curves start from the origin, $(q=0,~\omega=0)$.
The first branch $\omega_1(q)$ corresponds to the orthogonal (out-of-plane) bending vibrations
of the nanoribbon; the second branch $\omega_2(q)$ describes the bending planar (in-plane)
vibrations in the plane. These branches approach smoothly the axes $q$:
$\omega(q)/q\rightarrow 0$, when $q\rightarrow 0$, so that the corresponding long-wave phonons
possess no dispersion. The third $\omega_3(q)$ and fourth $\omega_4(q)$ branches
of the dispersion curves
correspond to the out-of-plane twisting and in-plane longitudinal vibrational modes, respectively.
The corresponding long-wave modes possess nonzero dispersion, so that we can define
two limiting values,
$$
v_t=h_x\lim_{q\rightarrow 0}\frac{\omega_3(q)}{q},~~\mbox{and}~~
v_l=h_x\lim_{q\rightarrow 0}\frac{\omega_4(q)}{q}
$$
which defines the sound speeds of twisting and longitudinal phonons in the nanoribbon.
The values of these speeds depend on the nanoribbon chirality. Nanoribbons of the same
width but different chiralities support phonons with different sound speeds: armchair
nanoribbon with $K=20$ (width $D_y=11.05$~\AA) has the velocities
$v_t=3400$~m/s and $v_l=24900$~m/s, whereas for the zigzag nanoribbon with $K=12$
(width $D_y=11.34$~\AA) these velocities are smaller,
$v_t=1900$~m/s and $v_l=13200$~m/s. When the width of the nanoribbon grows, the velocity of
the longitudinal sound remains unchanged while the velocity of the twisting modes
decreases monotonically.

\section{Edge modes}

Solution of the eigenvalue problem (\ref{f10}) demonstrates that the nanoribbon can support
vibrational modes localized at the edges, the so-called {\em edge phonons}. To find such
surface modes, we should consider the nanoribbons of a sufficiently large width, and
we take the large number of atoms in the elementary cell, $K=120$.

To analyze oscillations (\ref{f9}), we define the distribution function
of the oscillatory energy in the elementary cell of the nanoribbon as follows
$$
p_i=M_i(|e_{3(i-1)+1}|^2+|e_{3(i-1)+2}|^2+|e_{3(i-1)+3}|^2)/M_0\ge 0,
$$
where $i=1,2,...,K$ is the number of atoms in the cell, $M_i$ is the mass of the $i$-th
atoms, $e_j$ is the component of the eigenvector ${\bf e}=\{e_j\}_{j=1}^{3K}$.
The energy distribution is normalized by the following condition,
$$
\sum_{i=1}^K p_i=({\bf Me},{\bf e})/M_0=1.
$$
We introduce the parameter characterizing the transverse energy localization in the nanoribbon
as follows,
$$
d=1/\sum_{i=1}^{K}p_i^2.
$$
This parameter characterizes the inverse width of the energy localization
in the elementary cell. If the vibrational mode is localized only on one atom
(i.e. there exists an atom $i_0$ for which $p_{i_0}=1$), the inverse width is $d=1$.
In the opposite limit, when the vibrational energy is distributed
equally  on all atoms ($p_i\equiv 1/K$), we have $d=K$, so that in a general case
$1\le d\le K$.

\begin{figure}[t]
\includegraphics[angle=0, width=1.\linewidth]{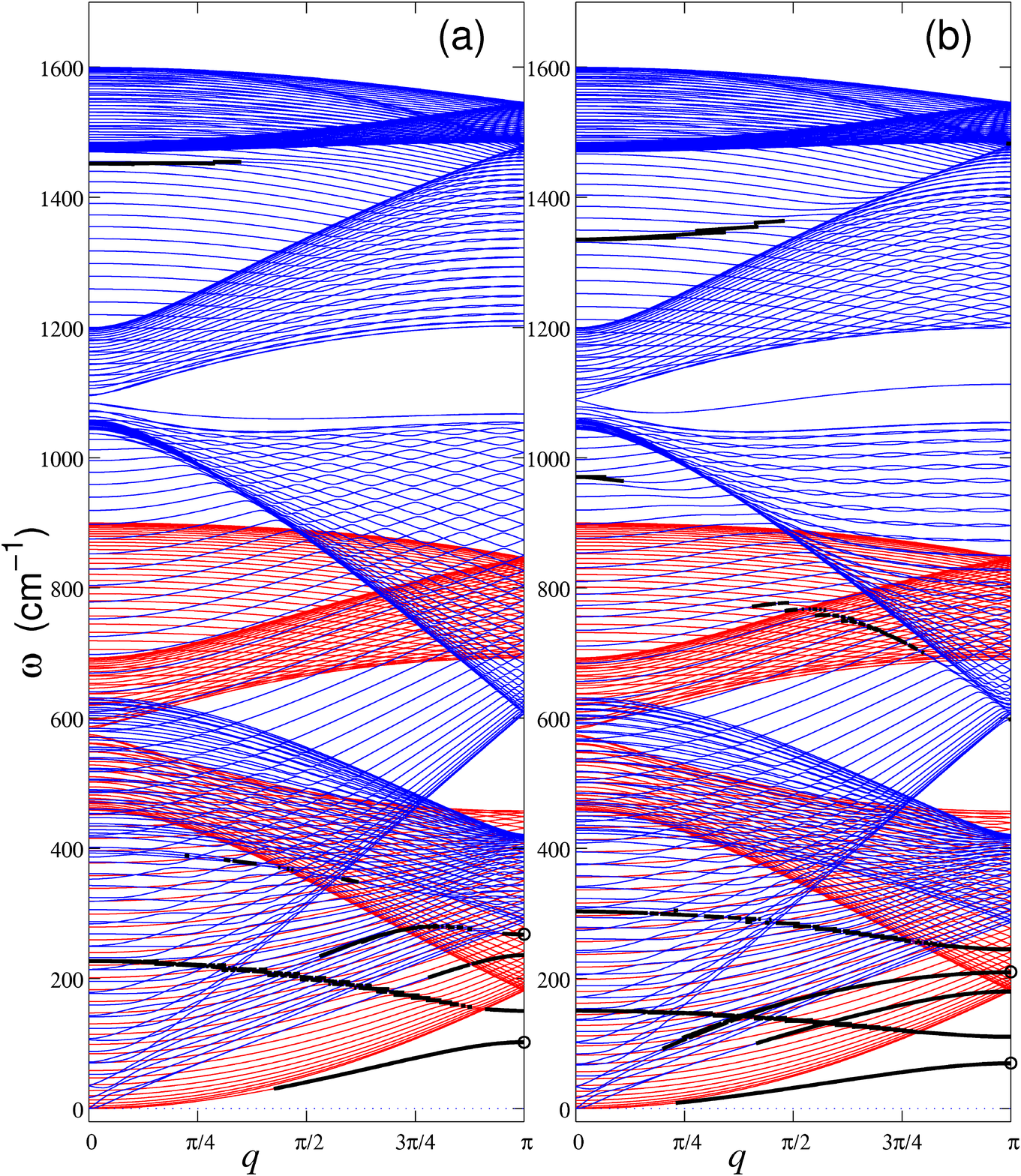}
\caption{  (color online)
Structure of $3K$ dispersion curves of the armchair nanoribbons for
(a) mass of the edge atoms  $M_1=13m_p$ and (b) $M_1=30m_p$.
The number of atoms in the elementary cell is $K=120$.
Blue (black) curves correspond to in-plane vibrations, red (gray) curves -- to the out-of-plane
vibrations, thick black curves -- oscillations localized at the nanoribbon edges.
The marked points show the band edges where nonlinear breathers can appear
in the form of the longitudinally localized edge modes.
}
\label{fg04}
\end{figure}

\begin{figure}[t]
\includegraphics[angle=0, width=1.\linewidth]{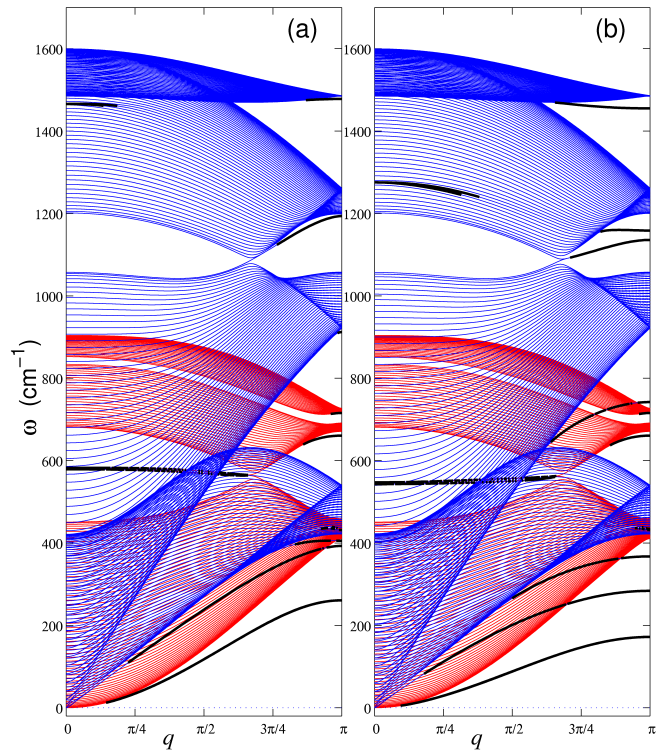}
\caption{  (color online)
Structure of $3K$ dispersion curves of the zigzag nanoribbons for
(a) mass of the edge atoms  $M_1=13m_p$ and (b) $M_1=30m_p$.
The number of atoms in the elementary cell is $K=120$.
Blue (black) curves correspond to in-plane vibrations, red (gray) curves -- to the out-of-plane
vibrations, thick black curves -- oscillations localized at the nanoribbon edges.
}
\label{fg05}
\end{figure}

To be more specific, we define the vibrational mode as being localized provided  $d<20$.
Our numerical analysis of the oscillatory eigenmodes shows that localization may occur
only at the edges of the nanoribbon. The structure of the $3K$ dispersion curves
of the armchair and zigzag nanoribbons are shown in Figs.~\ref{fg04} and \ref{fg05},
where thick lines mark the edge modes of the nanoribbon.
Such localized surface modes exist in both armchair and zigzag nanoribbons.
\begin{table*}[t]
\caption{
Frequencies $\omega$ (dimension $[\omega]=$cm$^{-1}$) of the localized edge oscillations
with the wave numbers $q=0$ and $q=\pi$, depending on the nanoribbon chirality and mass
of the edge atom $M_1$ (dimension $[M_1]=m_p$). Out-of-plane vibrations are marked
by italic. Oscillations with the wave number $q=0$ are underlined.
Corresponding energy densities of the edge modes are presented
in Figs.~\ref{fg06}  and \ref{fg07}.
}
\label{tb1}
\begin{tabular}{cccccccccccccccc}
 type & ~~$M_1$~~ & 1 & 2 & 3 & 4 & 5 & 6 & 7 & 8 & 9 & 10 & 11 & 12 & 13 & 14\\
\hline
armchair& 13 &\underline{\emph{227}}&\underline{1452}&\emph{102}&\emph{150}&236&268&1067&- &- &- &- &- &- &- \\
armchair& 30 &\underline{\emph{151}}&\underline{303}&\underline{971}&\underline{1335}&\emph{70}&\emph{110}&180&180&210&245&598&\emph{690}&1113&1482\\
zigzag& 13 &\underline{\emph{583}}&\underline{1466}&\emph{261}&393&406&\emph{431}&\emph{432}&\emph{661}&\emph{716}&912&1194&1478&-&-\\
zigzag& 30 &\underline{\emph{547}}&\underline{1276}&\emph{172}&284&368&\emph{413}&\emph{432}&\emph{661}&\emph{716}&742&1136&1159&1455&-\\
\hline
\end{tabular}
\end{table*}

Now we consider in more details the edge modes with the wave numbers $q=0$ and $q=\pi$.
The values of the frequencies of these modes are summarized in Table~\ref{tb1}. When the
mass of the edge atom is $M_1=13m_p$ (the effective edge atom corresponds to the
molecular group CH), the armchair nanoribbon has 7 edge modes (3 out-of-plane and 4 in-plane),
and zigzag nanoribbon has 12 edge modes (6 out-of-plane and 6 in-plane). If the mass of
the edge atoms is increased, the number of localized modes growths as well.
For example, for $M_1=30m_p$ (when the effective edge atom corresponds to the molecular
group CHOH), the armchair has already 14 edge modes (4 out-of-plane and 10 in-plane),
and the zigzag nanoribbon has 13 edge modes (6 out-of-plane and 7 in-plane).

We notice that for $K=120$, an increase of the mass of the edge atoms does not change
substantially the dispersion curves because the number of the edge atoms is much smaller
than the number of atoms in the elementary cell (elementary cell of the armchair
nanoribbon has 4 edge atoms, and the zigzag nanoribbon -- 2 edge atoms). However,
the surface modes are localized at the edge atoms, so that a change
of the mass of the edge atoms change the mode frequency substantially. As can be seen
from Figs.~{\ref{fg04}} and {\ref{fg05}}, a change of the mass of the edge atoms from
$M_1=13m_p$ to $M_1=30m_p$ leads mainly to a shift of the dispersion curves corresponding
to the edge modes. This behavior provides an additional indication of the existence of
strongly localized surface modes with the properties different
from those of the bulk modes.

\begin{figure}[t]
\includegraphics[angle=0, width=.8\linewidth]{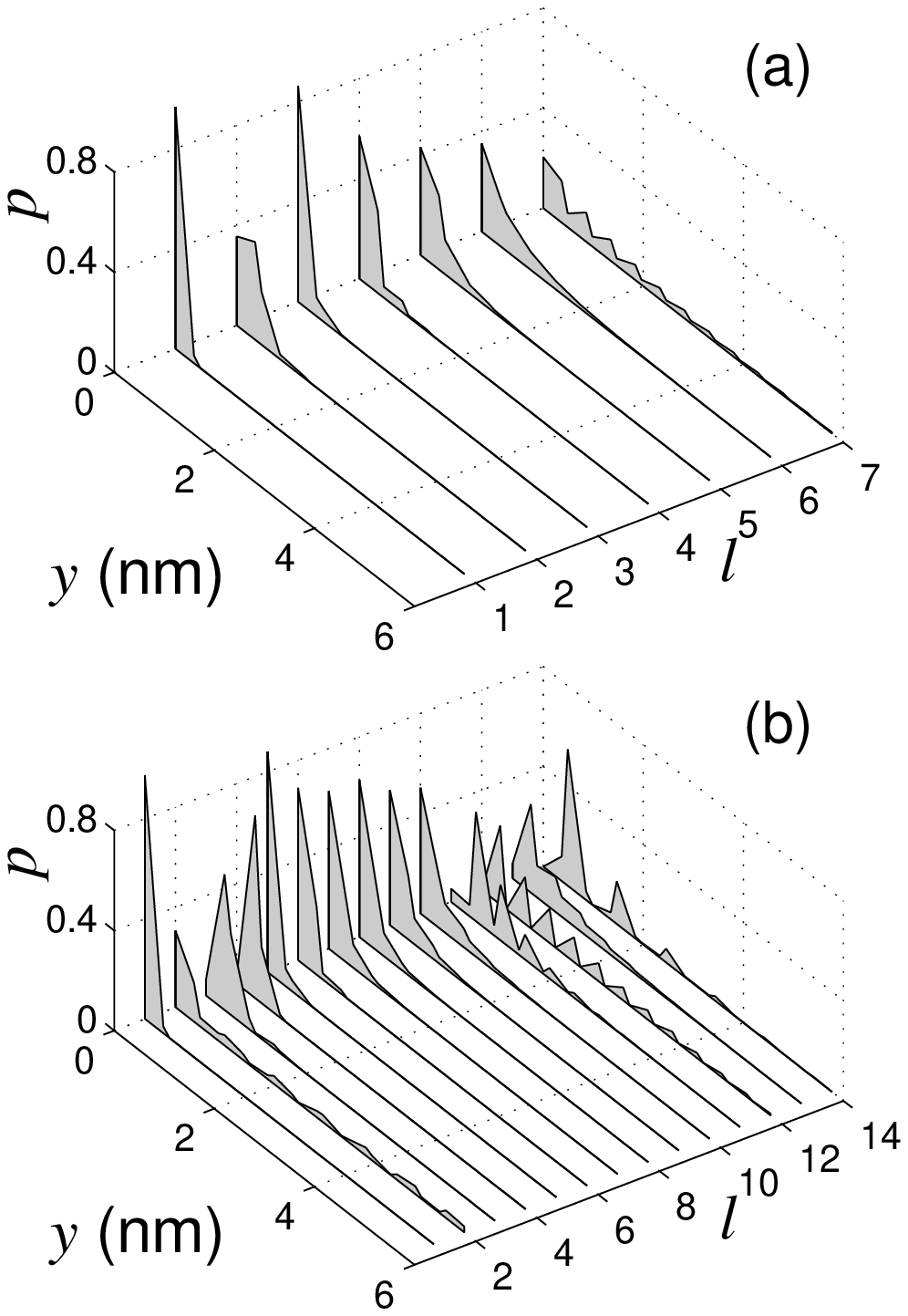}
\caption{
Transverse distribution of energy of the edge modes in the armchair nanoribbon with
the wave numbers $q=0$, and $q=\pi$ for the mass of the edge atoms: (a) $M_1=13m_p$
($q=0$ for $l\le2$, $q=\pi$ for $l\ge3$) and (b) $M_1=30m_p$
($q=0$ for $l\le4$, $q=\pi$ for $l\ge5$).
Here $y$ is the distance to the edge of the nanoribbon, $l$ is the mode number,
$p(y)$ is the energy density
normalized by the condition: $\int_0^\infty p(y)dy=1$.
The values of the frequencies are summarized in Table~\ref{tb1}.
}
\label{fg06}
\end{figure}

\begin{figure}[t]
\includegraphics[angle=0, width=.8\linewidth]{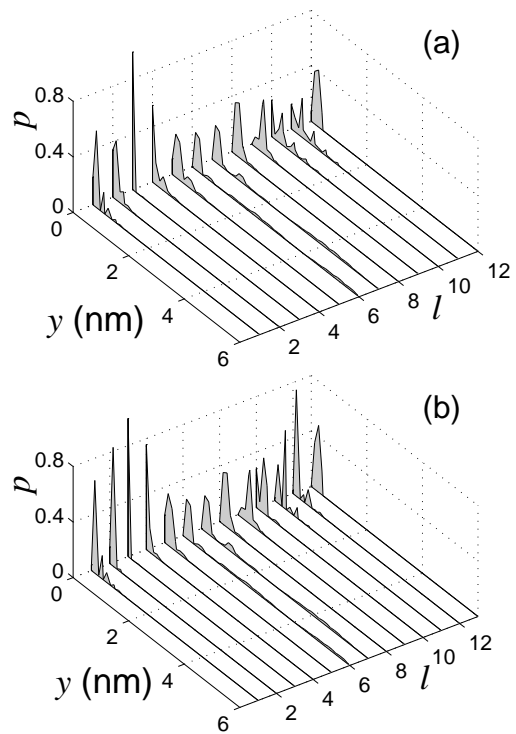}
\caption{
Transverse distribution of energy of the edge modes in the zigzag nanoribbon with
the wave numbers $q=0$, and $q=\pi$ for the mass of the edge atoms: (a) $M_1=13m_p$
and (b) $M_1=30m_p$ ($q=0$ for $l\le2$, $q=\pi$ for $l\ge3$).
Here $y$ is the distance to the edge of the nanoribbon, $l$ is the mode number,
$p(y)$ is the energy density
normalized by the condition: $\int_0^\infty p(y)dy=1$.
The values of the frequencies are summarized in Table~\ref{tb1}.
}
\label{fg07}
\end{figure}

Transverse distribution of the energy of the surface localized states in the graphene
nanoribbon is shown in Figs.~\ref{fg06} and \ref{fg07}.  As follows from these results,
the oscillations are mainly localized near the nanoribbon edges. For the armchair nanoribbon,
the oscillations penetrate inside the stripe for less than 4 nm, and for the zigzag nanoribbon
the oscillations decay for no more than 2 nm. For $M_1=13m_p$, the edge atoms of the
armchair nanoribbon always participate in the surface vibration, whereas in zigzag
nanoribbons the edge atoms may not contribute into the oscillatory mode, as shown
in Fig.~\ref{fg07} for the modes with the numbers $l=6$, 7, 8 9).

\begin{figure*}[t]
\begin{center}
\includegraphics[angle=0, width=.9\textwidth]{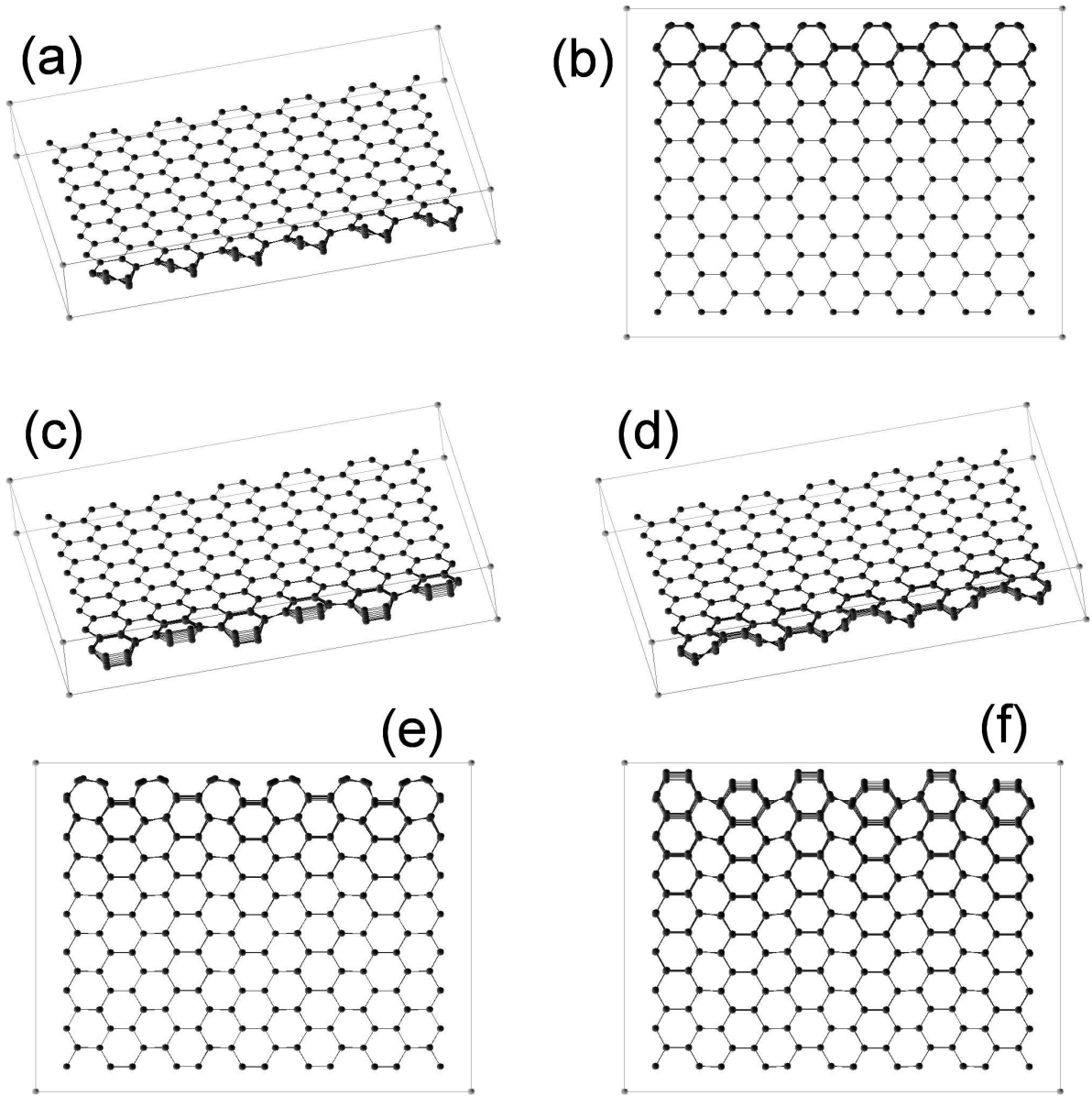}
\end{center}
\caption{\label{fg08}\protect
Six edge modes of the armchair nanoribbon. Shown are the displacements of atoms
of the nanoribbon at quarter of an oscillation period, for:
(a) out-of-plane vibration with $\omega=227$~cm$^{-1}$, $q=0$;
(b) in-plane vibration with $\omega=1452$~cm$^{-1}$, $q=0$;
(c) out-of-plane vibration with $\omega=102$~cm$^{-1}$, $q=\pi$;
(d) out-of-plane vibration with $\omega=150$~cm$^{-1}$, $q=\pi$;
(e) in-plane vibration with $\omega=236$~cm$^{-1}$, $q=\pi$;
(f) in-plane vibration with $\omega=268$~cm$^{-1}$, $q=\pi$.
Mass of the edge atom is $M_1=13m_p$.
}
\end{figure*}

\begin{figure*}[tbhp]
\begin{center}
\includegraphics[angle=0, width=0.9\textwidth]{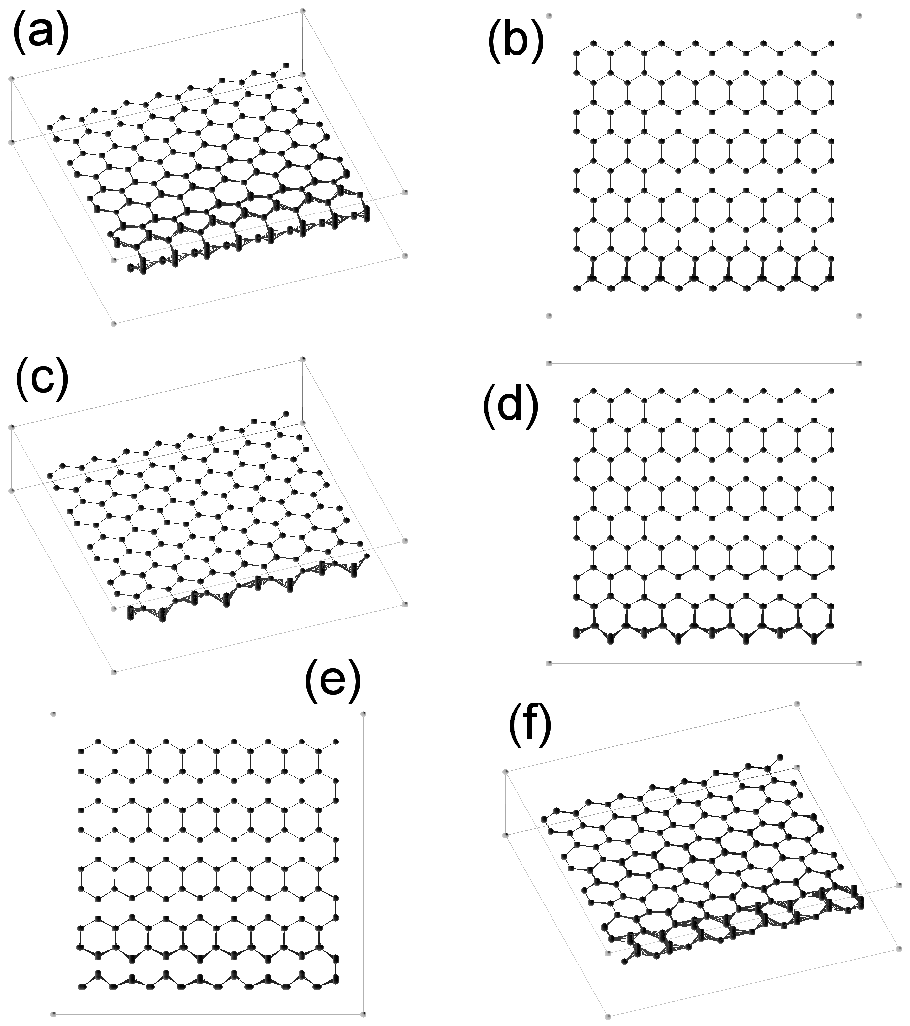}
\end{center}
\caption{\label{fg09}\protect
Six edge modes of the zigzag nanoribbon. Shown are the displacements of atoms
of the nanoribbon at quarter of an oscillation period, for:
(a) out-of-plane vibration with $\omega=583$~cm$^{-1}$, $q=0$;
(b) in-plane vibration with $\omega=1466$~cm$^{-1}$, $q=0$;
(c) out-of-plane vibration with $\omega=261$~cm$^{-1}$, $q=\pi$;
(d) in-plane vibration with $\omega=393$~cm$^{-1}$, $q=\pi$;
(e) in-plane vibration with $\omega=406$~cm$^{-1}$, $q=\pi$;
(f) out-of-plane vibration with $\omega=431$~cm$^{-1}$, $q=\pi$.
Mass of the edge atoms is $M_1=13m_p$.
}
\end{figure*}

The specific structure of the edge oscillations in the graphene nanoribbon
are shown in Figs.~\ref{fg08}
and \ref{fg09}. These edge modes can be divided into three groups. The first group is
composed by low-frequency out-of-plane vibrations which mainly occur due to deformation
of the soft dihedral angles. For the armchair nanoribbon with the mass of the edge atoms
$M_1=13m_p$, these oscillations have the frequencies $\omega=102$, 150 and 227~cm$^{-1}$,
whereas for the zigzag nanoribbons the similar modes have the frequencies:
$\omega=261$, 431, 432, 583, 661, and 716~cm$^{-1}$.
As can be seen from Figs.~\ref{fg06}(a), \ref{fg07}(a), and Table~\ref{tb1},
these oscillations are the most localized.
The main vibrational energy is concentrated in the edge atoms, which oscillate
orthogonally to the nanoribbon plane, except the modes in the zigzag nanoribbon
with the frequencies 431, 432, 661, and 716~cm$^{-1}$ (here the edge atoms are static
while the hexagons which contains these atoms rotate).

The second group of the edge modes is composed by the edge in-plane vibrations which
occur due to a deformation of the valent angles. For the armchair nanoribbon
these modes have the frequencies 236 and 268~cm$^{-1}$, and for the zigzag nanoribbon,
the frequencies 393 and 406~cm$^{-1}$.

Finally, the third group of the surface modes is composed by high-frequency edge
oscillations which occur in the nanoribbon plane which occur due to the deformation
of the hard valent bonds. For the armchair nanoribbons, these are the edge modes
with the frequencies 1067 and 1452~cm$^{-1}$, and for the zigzag nanoribbon,
with the frequencies 912, 1194, 1466, and 1478~cm$^{-1}$.

We would like to mention that in the paper~\cite{p4} the edge vibrations were studied
by another method by using the second-generation reactive empirical bond order (REBOII)
potential and density-functional theory calculations. This method does not allow to study
nanoribbons of large width so that the authors of Ref.~\cite{p4} considered the
nanoribbons with $K=20$ atoms of hydrogen in the elementary cell, taking into account
only the oscillations of the C--H groups. They found several types of
the out-of-plane edge vibrational modes: two edge modes for the armchair nanoribbon
(with the frequencies 365 and 858~cm$^{-1}$), and one mode for the zigzag nanoribbon
(with the frequency 610~cm$^{-1}$).

In our work, using the mechanical model and direct molecular-dynamics numerical
simulations allow to study much wider nanoribbons and, therefore, to find many more
types of surface modes. We consider the edge \mbox{C--H} atomic group
as a single edge atom with the mass $M_1=13m_p$, so that we are not able to study the
vibrational mode associated with the motion of one hydrogen atom with the frequency
858~cm$^{-1}$ from the paper~\cite{p4}. However, for other two
vibrational modes with the frequencies  365 and 610~cm$^{-1}$,  both carbon and hydrogen
atoms participate in vibrations, so that we can find similar modes by applying our approach,
with the frequencies 227 and 583~cm$^{-1}$.  Taking into account the motion of the
hydrogen atoms leads to a correction of the oscillation  frequency, as well as to
a larger number of the edge modes, due to additional modes induced by separate motion
of hydrogen and carbon atoms but not the whole group C--H.

\section{Longitudinal localization and surface solitons}

The surface oscillations described above are localized at the edges of the stripe, and they
correspond to quasi-one-dimensional excitations of the nanoribbon.  Therefore, similar to other
types of nonlinear modes, we may expect that such surface mode may experience the longitudinal
localization with the formation of breather solitons. Such novel type of surface
breathers exist due to the transverse energy localization in the surface state and
nonlinearity-induced longitudinal localization along the nanoribbon edge.

To analyze the existence of breather solitons, we consider the nanoribbon of a fixed length
and free edges with $N=500$ periods.  We introduce friction at the far edges in order to absorb
phonons propagating away from the localization region. To find spatially localized
nonlinear states,
we integrate numerically the system of equations (\ref{f7}) ($1\le n\le N$) with the initial
conditions corresponding to a localized state at the edge of the nanoribbon,
\begin{equation}
{\bf x}_n(t)=A_n{\bf e}\exp(iqn-i\omega t),
\label{f11}
\end{equation}
where $A_n=A\;{\rm sech}\{\mu[(n-N/2)+1/2]\}$, with the parameter $\mu>0$
characterizing the inverse width
of the initial excitation, and the amplitude $A>0$. A complex unit $3M$-dimensional vector
${\bf e}={\bf e}_r+i{\bf e}_i$, a solution of the linear equation (\ref{f10}),
corresponds to the edge mode with the wave numbers $q=0,~\pi$ and the frequency $\omega$.

For the system of equations (\ref{f7}), the ansatz (\ref{f11})
corresponds to the following initial conditions,
\begin{eqnarray}
\label{f12}
{\bf x}_n(0)&=&A_n[{\bf e}_r\cos(q\tilde{n})-{\bf e}_i\sin(q\tilde{n})],\\
\dot{\bf x}_n(0)&=& A_n\omega[{\bf e}_r\sin(q\tilde{n})-{\bf e}_i\cos(q\tilde{n})],
\nonumber
\end{eqnarray}
where $\tilde{n}=n-N/2+1/2$.

The breather frequency can only split-off from one of the linear dispersion bands.
As follows from Figs.~\ref{fg04} and \ref{fg05}, the frequency branches corresponding
to the edge vibrations have either minimal or maximum values at $q=0$ and $q=\pi$,
so that we use these values in the
initial conditions (\ref{f12}).

By changing the value of the initial amplitude $A$ and the inverse width $\mu$, and
analyzing the dynamics of the system (\ref{f7}) with the initial conditions (\ref{f12}),
we may arrive at the conclusion if the system under consideration can support
spatially localized states in the form of the breather solitons, as self-trapped
nonlinear modes splitting-off from the corresponding edge of the linear dispersion band.
If such a breather exists, the initial excitation (\ref{f12}) will experience no
substantial spreading, whereas it will spread out if the breather conditions are not satisfied.

\begin{figure}[t]
\begin{center}
\includegraphics[angle=0, width=1\linewidth]{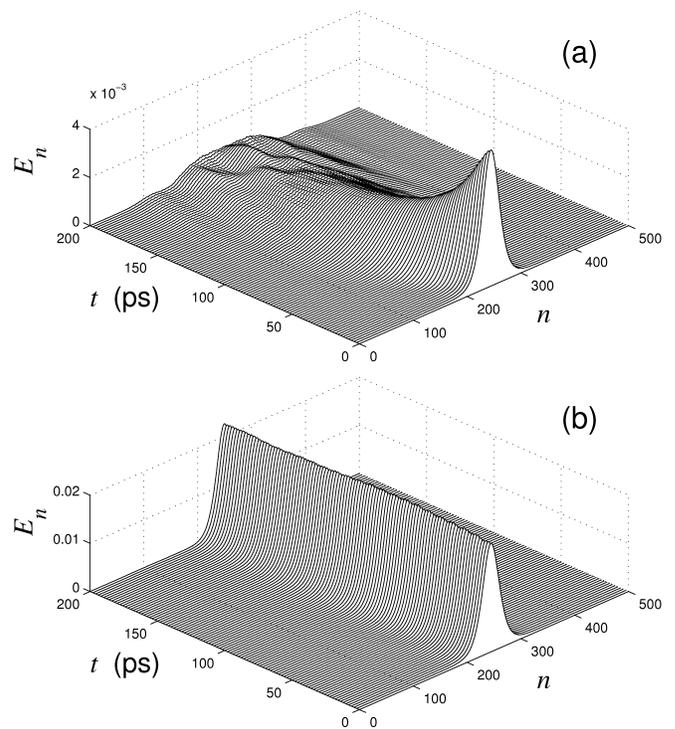}
\end{center}
\caption{\label{fg10}\protect
Dynamics of the initially localized edge excitation of the
armchair nanoribbon (wave number $q=\pi$, frequency $\omega=268$~cm$^{-1}$, and
inverse dimensionless width $\mu=0.05$) for the initial amplitude:
(a) $A=0.004$ and (b) $A=0.008$~\AA.  Shown is the temporal evolution
of the the kinetic energy $E_n$ in two first edge layers of atoms of the nanoribbon
($k=1$, 2, 3, 4).  Mass of the edge atoms is $M_1=13m_p$.
}
\end{figure}

Our extensive numerical simulations reveal that the edge breather are not possible
for the zigzag nanoribbons, and the initial excitation of the form (\ref{f12}) always
spreads. However, for the armchair nanoribbons
the longitudinal localization of the edge modes becomes possible with the wave number
$q=\pi$ and frequency $\omega=102$ (out-of-plane vibration) and 268~cm$^{-1}$ (in-plane vibration),
 see Fig.~\ref{fg10}. For the inverse width $\mu=0.05$ and amplitude
$A=0.008$~\AA, the initial localized vibrational state with the frequency
$\omega=102$~cm$^{-1}$ decays slowly, but this decay stops when the initial amplitude
is increased to the value $A=0.016$~\AA, and it evolves into a localized breathing mode with
the breather frequency 103~cm$^{-1}$.  When we increase the initial amplitude to the value
$A=0.032$~\AA, we observe the generation of two breathers which interact nonlinearly.

Similarly, for the planar edge vibration with the frequency $\omega=268$~cm$^{-1}$
the initial localized state spreads when its amplitude is $A=0.004$~\AA~ [see Fig.~\ref{fg10}(a)],
it gets self-trapped for $A=0.008$~\AA~ and generates a breather with the frequency 267~cm$^{-1}$
[see Fig.~\ref{fg10}(b)], and for $A=0.016$~\AA~ it creates two interacting breathers.

Our studies demonstrate that armchair nanoribbons can support two types of spatially
localized surface breathers.
One type of breathers originates from the localization of out-of-plane edge mode with the frequency
$\omega=102$~cm$^{-1}$. This edge mode is shown in Fig.~\ref{fg08}(c), where the edge atoms
of the neighboring elementary cells oscillate with the opposite phases in the direction perpendicular
to the nanoribbon plane. The other type of breathers originates from the longitudinal
localization of the in-plane edge mode with the frequency $\omega=268$~cm$^{-1}$.
This edge mode is shown in Fig.~\ref{fg08}(f), where
the edge atoms oscillate transversally in the nanoribbon plane. The frequency of
the former breather grows with the amplitude, while the frequency of the latter breather
decays with the amplitude, in accord with
the structure of the dispersion curves presented in Fig.~\ref{fg04}.

\section{Conclusions}

We have demonstrated that both armchair and zigzag graphene nanoribbons can support
various types of vibrational Tamm states in the form of phonons localized at the edges
of the graphene stripe. We have shown that such localized modes appear in the gaps of the
frequency spectrum of the nanoribbon oscillations, and they can be classified as in-plane and
out-of-plane surface modes. In the former case, all atoms oscillate in the plane
of the nanoribbon, whereas in the latter case the atoms oscillate perpendicular
to the plane. We have found that some of these edge modes can experience self-localization
along the nanoribbon edge with the formation of spatially localized breather solitons.
Our molecular dynamics simulations have revealed the existence of two types of such breather solitons,
out-of-plane vibrational breathers with the frequency 103~cm$^{-1}$ and in-plane
breathers with the frequency 267~cm$^{-1}$.

\section*{Acknowledgements}

Alex Savin acknowledges hospitality of the Nonlinear Physics Center at the Australian
National University. This work was supported by the Australian Research Council.
The authors also thank the Joint Supercomputer Center of the Russian Academy of Sciences
for using their computer facilities.

\end{document}